\documentclass[usenatbib,usegraphicx]{mn2e}
\usepackage{psfig,epsf}

\newcommand{\kms}{\ensuremath{ \mbox{km}\ \mbox{s}^{-1}}}

\newcommand{\HI}{H\,{\sc i}\ }

\newcommand{\mnras}{MNRAS}
\newcommand{\apj}{ApJ}
\newcommand{\apjs}{ApJS}
\newcommand{\aj}{AJ}
\newcommand{\aap}{A\&A}

\newcommand{\araa}{ARA\&A}

\newcommand{\pasp}{PASP}
\newcommand{\msunyr}{\ensuremath{M_{\odot}\ {\rm yr}^{-1}}}
\begin{document}
\title[Young stars in NGC 6822.]{Young Stars in the Outer H\,{\sc i} Disc of NGC 6822}
\author[de Blok \& Walter]{W.J.G.~de~Blok$^1$, F.~Walter$^2$\thanks{Jansky Fellow}\\
$^1$ Department of Physics and Astronomy, Cardiff University, 5 The Parade, Cardiff CF24 3YB, United Kingdom\\
$^2$ NRAO AOC,  P.O. Box O,  1003 Lopezville Road, Socorro, NM 87801-0387, USA}
\maketitle
 
\begin{abstract} 

We present wide-field optical imaging covering the entire neutral
hydrogen disc of the Local Group dwarf galaxy NGC 6822. These
observations reveal the presence of numerous blue, young stars at
large galactocentric radii well beyond $R_{25}$. Blue stars are also
found that are associated with NGC 6822's companion \HI cloud,
indicating that star formation was triggered in the companion in the
last $10^8$ yr.  In general, blue stars are present where the \HI
surface densities reach values $ \ga 5 \times10^{20}$
cm$^{-2}$. However, over one-third of the blue stars detected within
the \HI disk are found at lower surface densities.
The young stars trace the distribution of the neutral hydrogen in the
inner disk, but seem to be avoiding the supergiant
\HI shell in NGC 6822, setting a lower limit for its age of $10^8$ yr.
The extended distribution of young stars implies that stars can form
at large galactocentric radii in dwarf galaxies; the \HI is therefore
not necessarily much more extended than the stellar population.  This
finding has important consequences for the chemical enrichment of the
interstellar medium throughout (dwarf) galaxies.
\end{abstract}


\section{Introduction}

NGC 6822 is the most nearby dwarf irregular galaxy outside the Milky
Way/LMC/SMC system and is therefore a prime target to study in detail
the interplay of star formation and the atomic Interstellar Medium
(ISM). At a distance of only 490 kpc, neutral hydrogen observations
reach an unprecedented physical resolution and stellar population
studies can be performed even with ground-based telescopes.  

Extensive studies of the distribution and evolution of the stellar
populations in NGC 6822 were done by e.g.,
\citet{hodge77,hodge80}, \citet{hodge91} and 
\citet{gallart1,gallart2,gallart3}. These studies find a star formation 
history which is stochastic, low-level, but on 
average constant,
with an increase in star formation rate (SFR) of a factor of a few around
$\sim 100$ Myr ago, which persists until the present.
Various estimates for the recent SFR
\citep{israel,gallart3,hodgeHA2} all indicate a low value between
$\sim 0.02$ and $\sim 0.07\ \msunyr$, with an average of 0.06 \msunyr\
\citep{mateo98}.
NGC 6822 is a metal poor galaxy, with an ISM metal abundance of $\sim$
0.2-0.3 $Z_{\odot}$ \citep{pagel,evan89}. Similar values are found for
the stellar abundances \citep{venn}.
The intermediate-age stellar population, as traced by C and M stars,
was mapped by \citet{letarte}, who found that these are distributed in
an elliptical halo that does neither trace the distribution of the
H\,{\sc i}, nor that of the main optical component. A number of C
stars are found outside the \HI disc, though this may be a common
phenomenon for dwarf galaxies (e.g.\ \citealt{battinelli}).

The most recent observations in the 21-cm line of neutral hydrogen are
presented in \citet{dBW2000} (hereafter dBW00) and \citet{weldrake}.
These show that the galaxy is dark matter-dominatated, gas-rich, and
reveal many interesting features, such as a highly structured \HI
disc, a super-giant \HI shell, and the presence of a companion
cloud.
In this Letter, we discuss the properties and distribution of the
stellar population, with emphasis on the young
blue stars.  

\section{Observations and Analysis}
NGC 6822 was observed with the 40$''$ telescope at Siding Spring
Observatory, Australia from 5-12 July 1999. We used the Imager with a
2048$\times$2048 Tek CCD with a pixel scale of 0.6$''$ pixel$^{-1}$,
resulting in an effective field of view of $\sim 20' \times 20'$.  The
entire extent of the \HI disc of NGC 6822 was covered in 8 pointings
in $B$ and $R$. Each field was observed for 2500s in $R$ and 3000s in
$B$. The total area covered was $\sim 2800$ square arcmin. Standard
stars were observed throughout the run. The seeing was fairly constant
at 1.5$''$--2$''$. Conditions were photometric for 3 nights and
variable during the rest of the run. We used the overlap regions
between fields to put non-photometric images on the same flux
scale. We found that due to this procedure absolute photometry is
accurate to  $\sim 0.1$ mag.  We used the DOPHOT \citep{dophot}
crowded field photometry package to create a catalog of stellar
objects in $B$ and $R$ in the observed area. Best results were
obtained by using an analytical point spread function, adjusted to the
typical seeing size for each field, and using a $15 \times 15$ pixels
box-averaged median background.  Stellar detections in both bands were
matched by cross-correlating the two lists, assuming a $1.5''$
uncertainty in the positions. This procedure yielded a catalogue of
$\sim 28,000$ stellar objects, the great majority of them being
Galactic foreground stars.

\section{Blue Stars in NGC 6822}

\begin{figure}
\hfil\psfig{figure=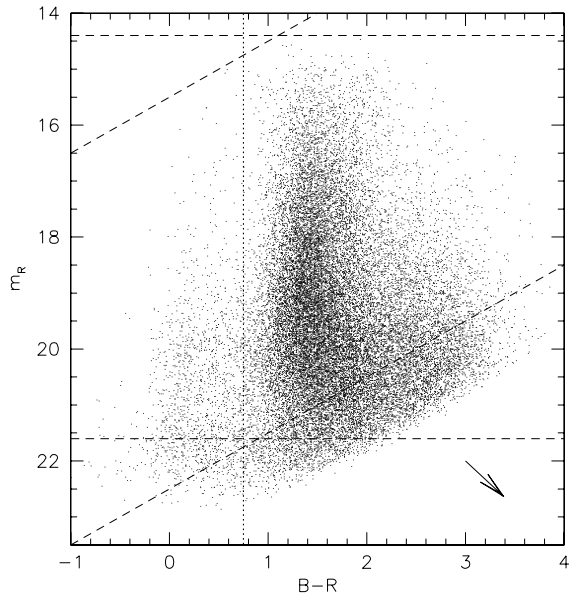,width=\hsize}\hfil
\caption{Colour magnitude diagram in $B$ and $R$. The
lines at $m_R = 14.4\ (21.6)$ indicate $R$-band completeness limits for
bright (faint) objects.  The diagonal lines indicate bright
(faint) $B$-band limits of $m_B = 15.5\ (22.5)$. The vertical dotted
line indicates $B\!-\!R=0.75$ which is our  dividing
line for blue stars. The arrow indicates the extinction
and reddening vector appropriate for NGC 6822. \label{colmag} }
\end{figure}

\subsection{Colour-magnitude diagram}
In Fig.~\ref{colmag} the colour-magnitude diagram (CMD) of the entire
observed field in $B$ and $R$ is shown.  Two features are obvious: the
large concentration of stars at $B\!-\!R \sim 1.5$ and redder, and the
plume with bluer colors at $B\!-\!R \sim 0.4$. Stars with $B\!-\!R \sim 1.5$
are mainly Galactic foreground stars, though it is expected that some
bright red stars in NGC 6822 also contribute. Stars with $B\!-\!R \sim
0.4$ are the blue stars which (mostly) belong to NGC 6822.

Completeness limits at the faint magnitude end were derived by
plotting the histograms of the apparent $B$ and $R$ magnitudes. These
showed the expected steep increase of number of stars toward fainter
apparent magnitudes, followed by a sharp drop-off due to decreasing
completeness. The peaks in the histograms occurred at $m_R = 21.6$
and $m_B = 22.5$, and we take these as our fiducial completeness
limits.

We select the blue stars in the CMD by choosing only those detections with
$B\!-\!R<0.75$. 
This blue stellar population will be discussed in detail
in this Letter.  Using a distance modulus of $23.49 \pm 0.05$
\citep{gallart1}, we find that the colours and luminosities of the
blue stars associated with NGC 6822 are consistent with main sequence
stars with spectral types between B5 $(M_B \sim -1.3)$ and O5 $(M_B
\sim -5.8)$. There are possibly a few stars of even earlier type
present. The main sequence life-times of these stars vary from $\sim
80$ Myr (B5) to $\sim 1$ Myr or less (O5). Stars of later spectral
types must obviously be present in NGC 6822 as well, but cannot be
detected using the current data set.  The total observed magnitude of
the 1549 detected blue stars is $m_B = 12.0$ or $M_B = -12.5$ after
correction for extinction and distance (the total blue luminosity of
NGC 6822 is $M_B = -15.8$).

\subsection{Distribution and significance of Blue Stars}
\begin{figure*}
\hfil\psfig{figure=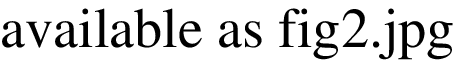,width=0.3\hsize}\hfil

\caption{Distribution of stars in NGC 6822. Dotted lines indicate 
the area surveyed.  The outer ellipse indicates $D_{25}$ ($15.5'\times
13.5'$ [RC3]), the inner ellipse the optical bar.  {\emph a)}
Distribution of blue stars. Grey dots: stars with $B\!-\!R<0.75$ and
$m_R > 19$; black dots: stars with $B\!-\!R<0.75$ and $m_R \leq
19$. {\emph b)} Distribution of blue stars overlaid on \HI surface
density map.  Contours indicate $\sigma = (2.5; 5) \cdot 10^{20}$
cm$^{-2}$ (inclination corrected).  For the stars no distinction in
apparent brightness is made here.  {\emph c)} As b), showing the
distribution of C stars from \citet{letarte}.  {\emph d)} Overlay of
blue stars on velocity dispersion map.  White: dispersion $\delta < 8$
\kms; light-grey: $8 \leq \delta < 12\ \kms$; dark grey: $12 \leq
\delta < 16\ \kms$.
\label{mommap}}

\end{figure*}
Figure~\ref{mommap}a shows the position of the blue stars compared to
the approximate extent of the optical galaxy, where we distinguish
between stars with $m_R > 19.0$ (``faint'') (grey dots) and stars with $m_R \le
19.0$ (``bright'') (black dots).  Also shown (Fig.~\ref{mommap}b) are the
positions of the blue stars in relation to the distribution of the
atomic interstellar medium (ISM) in NGC 6822 (see dBW00 and
\citealt{weldrake}).  
It is immediately clear that
\emph{the blue stars cover a much larger area than what is usually
regarded as the optical component of NGC 6822}; \emph{the distribution
of the blue stars is as extended as that of the inner main \HI
disk}. Additionally stars seem present in the NW \HI cloud. The only
areas seemingly devoid of blue stars are the centre of the large hole
and the tidal arm in the SE.  Note that the latter does not imply
there are no stars present there. It simply means that stars in this
area must be fainter than $m_R \sim 21.6\ (M_R \sim -1.8)$ in order not
to be detected in the current data set. Deeper surveys may perhaps
show a faint population there.

In Fig.~\ref{mommap}c we  show the distribution of C stars, which
trace the intermediate age ($\sim$ 3--5 Gyr) population (from
\citealt{letarte}).  
\emph{The distribution of the young stars is in stark contrast with that of
the older stellar population.}  The latter is distributed in an
elliptical halo that does not trace the distribution of the H\,{\sc
i}, and shows no internal structure.  The young stars, on the other
hand, have a clumpy, almost filamentary distribution. Two of the large
concentrations coincide with the opposite ends of the bar-like optical
structure. There is a larger ratio of bright to faint blue stars
inside the optical bar than outside, presumably indicating more recent
star formation there. 

Blue star candidates are scattered throughout the field; however, most
of these are likely to be false candidates (faint galactic blue stars,
or, more probably, stars that are scattered into the blue bin due to
magnitude uncertainties).  Nevertheless, as presumably the
contribution of false candidates within the disk of NGC 6822 is as
important as outside the disk, it is illustrative to quantify the
significance of the detected overdensities in the presence of this
background of false detections.  To this end we have used the observed
distribution of detected objects to calculate the number density
distribution by counting the number of detections in 30$''$ by 30$''$
boxes. This size was chosen as a compromise between preserving spatial
resolution and getting sufficient numbers of detections in each
box. The size of the box does not affect the conclusions.  The
creation and detection of false candidates in the field is a random
proces, so the presence or absence of candidates in each box can be
regarded as a shot-noise process, and the surface density of the false
candidates will be approximately constant across the field.

\begin{figure}
\psfig{figure=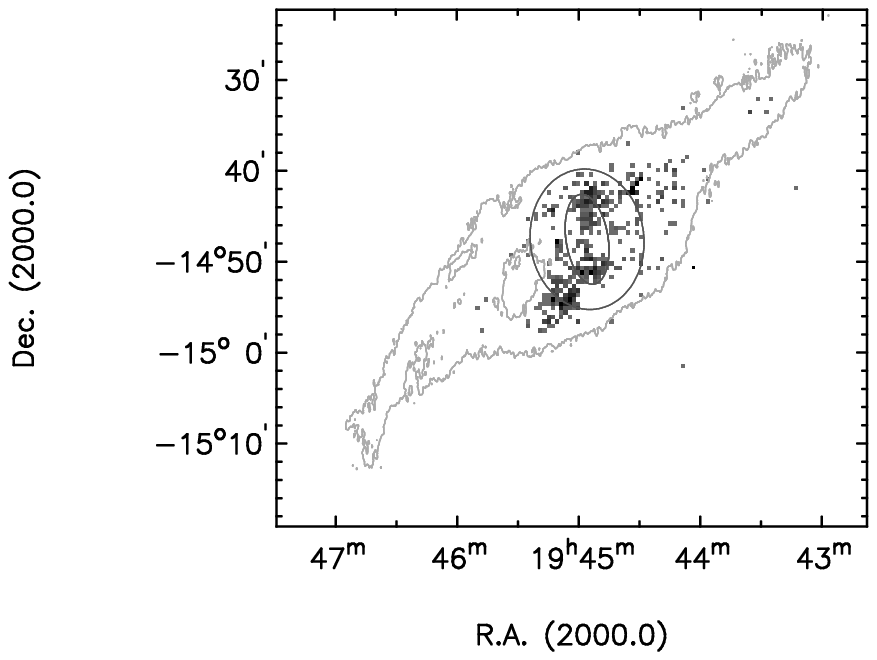,width=\hsize,clip=y}
\caption{The number density of blue stars averaged over $30'' \times 30''$ boxes. Only boxes containing 2 or more stars are shown. The chance that each of the boxes shown results from superpositions of false candidates is less than 0.1 percent. The correlation with the \HI distribution furthermore 
enhances their significance.
\label{densmap}}
\end{figure}

We have
investigated the number of false candidates in each box outside the $1
\times 10^{20}$ cm$^{-2}$ \HI column density contour (i.e., outside the
entire \HI disk as shown in Fig.~\ref{mommap}). We will refer to this
area as ``the field''.  There are 4955 30$''
\times 30''$ boxes in the field, of which 4868 have no false candidates, 84 have 1 false candidate and 3 have 2 false candidates. This distribution can be
well described by a Poisson distribution
$P(x)=(e^\lambda\lambda^x)/x!$, with $\lambda=0.018$. The cumulative
Poisson distribution then yields that the chance of getting 2 or more
false candidates in a $30'' \times 30''$ box due to shot-noise effects
is less than 0.1 percent. Figure~\ref{densmap} shows the number
density image, where we only show boxes with 2 or more candidates,
which are thus statistically very unlikely to have resulted from
random effects.  Note that the presence of the few significant boxes
outside the \HI disk is entirely consistent with the derived Poisson
distribution. Using this distribution we can also compute that we
expect $\sim 25$ false detections within the entire
\HI disk. This is in sharp contrast with the $\sim 1400$ stars
actually found.  A repeat of this analysis with different sized boxes
yields identical conclusions: the number density of false detections
can be described by a Poisson distribution and the features in the
stellar distribution as described above are statistically significant.

\begin{figure}
\psfig{figure=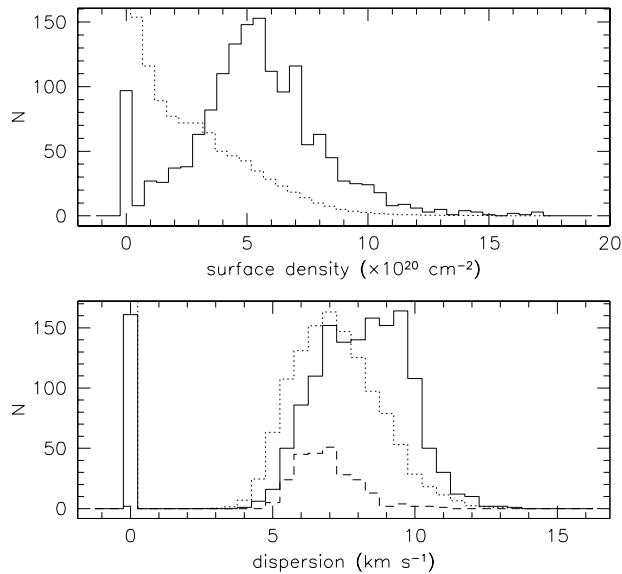,width=\hsize}
\caption{{\emph{Top:}} Histogram of inclination corrected 
\HI surface densities at $12\times40''$
($30\times100$ pc) resolution at the positions of the blue stars. The
peak at $\sigma=0$ is caused by possibly false stellar candidates
outside the
\HI disc of NGC 6822. The dotted line shows the arbitrarily scaled 
histogram of surface densities of the entire \HI disc. 
\emph{Bottom:} as top panel, but now showing the distribution of \HI velocity dispersions at the positions of the blue stars (solid line). The dashed histogram shows the dispersions of the \HI at the positions of the stars in the SE association (Sect.~3.3). The dotted line shows the arbitrarily scaled histogram of velocity dispersions of the entire \HI disc.
\label{coldensx}}
\end{figure}

Figure~\ref{coldensx} shows a histogram of the surface densities
$\sigma$ at the position of the blue stars (at $12'' \times 40''$ [$30
\times 100$ pc] resolution). Of the $\sim 1400$ stars detected in the
\HI disc, $\sim 900$ are found at surface densities of $\sigma \ga
5\times10^{20}$ cm$^{-2}$, close to or above the empirical
star-formation threshold (e.g.\ \citealt{skillman87, kenn89}).  It is
however worth noting that $\sim 500$ stars are found at surface
densities significantly lower than this star formation threshold
value. (These numbers do not include the presumed false candidates
outside the \HI disc). From the Poisson distributions derived above we
only expect $\sim 25$ false candidates in the \HI disk, clearly not
enough to explain the large number of stars at low surface
densities. For comparison, a normalized histogram showing the relative
distribution of surface densities of the entire \HI disc is
over-plotted in Fig.~\ref{coldensx}.

Figure \ref{mommap}d shows the positions of the blue stars
superimposed on a map of the velocity dispersion of the \HI gas.  A
histogram of the velocity dispersions at the positions of the blue
stars is shown in Fig.~\ref{coldensx}.  The average velocity
dispersion of the \HI gas at the positions of the blue stars is $\sim
8$ \kms, about 1 \kms\ higher than the average velocity dispersion of
the \HI gas in the entire disc.
Assuming an internal velocity dispersion of a few km s$^{-1}$ for a single
star cluster (or OB association) after its formation out of a
molecular cloud, we estimate that the observed young stars can travel
$\sim$ 100 pc away from their parent molecular cloud over their
lifetimes. It is thus unlikely that the young stars found at low
surface densities have traveled there from high-density regions.

\subsection{Details of the HI distribution}

{\bf The companion HI cloud.} A small concentration of OB stars at a
galactocentric distance of $\sim 5$ kpc is detected to the NW of the galaxy.  It
coincides with the position of the companion \HI cloud described in
dBW00. This companion is presumably interacting with NGC 6822 on
timescales of 100-200 Myr (see dBW00 for details).  The presence of
blue stars with ages less than $\sim$100 Myr gives additional support
to this scenario, the idea being that star formation was triggered in
the smaller companion as a result of the interaction.  Moreover, no
tracers for an intermediate age population of stars are found in the
\HI cloud, implying that we may indeed be seeing the first major
episode of star formation here.

{\bf The supergiant HI shell.}  It is striking that no large
population of bright blue stars is found inside the supergiant HI
shell (SGS).  In dBW00 we derived an age of $\sim 130$ Myr for
the SGS.  The blue stars trace ages up to $\sim80$ Myr so we can place
a lower limit on the age of the SGS of order $\sim100$ Myr. If the SGS
was `only' 130 Myr old, and if stars created the hole, we would expect
to detect at least some of the remnant stars.  The non-detection of
this population of stars implies that the SGS is probably older than
previously derived. As was noted by dBW00, it is difficult to derive
an accurate dynamical age for the SGS due to the lack of clear signs
of expansion.  As NGC 6822 rotates almost like a solid body throughout
the region of interest \citep{weldrake}, large coherent structures
such as the SGS can remain intact for many $10^8$ years, long after
the remnant stellar population has faded (see also \citealt{fabianelias}).

%The presence of C-stars, or an intermediate age population, in and
%near the hole may be a projection effect, depending on the geometry of
%the C-star halo, but if C-stars are indeed located in the hole, then
%this puts a (rather loose) upper limit on the age of the hole of $\sim
%3$ Gyr. Again, probing the older stellar population in more detail
%would help understand the possible evolution of the SGS.

{\bf The prominent concentration of young stars toward the
south-east.}  A prominent large concentration of blue stars outside
$D_{25}$ is readily seen to the south-east of the bar. It causes the
asymmetric appearance of NGC 6822 in the $B$-band (e.g.\
\citealt{hodge77}). It is in fact the most prominent association of blue
stars in the galaxy. While the velocity dispersion of the \HI
at the positions of the other large blue star concentrations hovers
between $\sim 8$ and $\sim 12$ \kms, at this position the average
dispersion is only $\sim 6$ to 7 \kms (cf.\ Figs.~\ref{mommap}d and
\ref{coldensx}).  There is a lack of prominent H$\alpha$ emission
in this region compared to the other concentrations (see
\citealt{hodgeHA}), suggesting that the SFR has declined in the past
10 Myr. The conspicuous lack of C stars indeed suggests that this region
has undergone its first major episode of star
formation some 10-100 Myr ago.  

\section{Conclusions}

We have presented a comparison of the distribution of blue stars and
the atomic ISM in NGC 6822. We find that the distribution of the blue
stars is almost as extended as that of the main \HI disk of NGC 6822.
Only the large hole and the tidal arms are devoid of blue stars.  Blue
stars are typically found at \HI surface densities $\ga 5 \cdot
10^{20}$ cm$^{-2}$, however over one-third of the blue stars is found
at lower surface densities. This could imply that {\it a)} stars can
form at surface densities below the canonical star formation
threshold; {\it b)} the ISM is very clumped at scales smaller than the
current resolution of $30 \times 100$ pc; {\it c)} star formation has
destroyed or ionized the regions of higher surface densities in which
these stars originally formed or {\it d)} the stars traveled from
higher density regions within their lifetimes.  However, due to their
short lifespan, blue stars are only expected to travel some 100 pc
from their places of birth and stellar cluster dispersion can not
explain the similarity of the distributions of the blue stars and the
atomic ISM.

Given the rather symmetrical distribution of the C-stars, it is clear
that neither the SGS, nor the interaction with the NW cloud has
affected the distribution of the old population (and therefore the
bulk of the mass in stars and stellar remnants). The processes that
are so dramatically visible in the H{\sc i}, must have had hardly any
impact on the overall mass distribution in NGC 6822. The presence of
blue stars over the main body of NGC 6822, as well as in the NW cloud
implies that star formation is occurring not just in the optical disk,
but that pollution of the ambient ISM is taking place on a larger
scale (similar conclusions can be drawn for massive galaxies as well, see e.g., \citealt{ronallen}). This may partly explain the flat abundance patterns found in
chemical abundances studies of low-mass galaxies (e.g.,
\citealt{kobulnicky}). 
No large population of young blue stars has been detected in the
supergiant \HI shell, setting a lower limit on its age of $\ga 100$
Myr (again, this does not rule out the presence of a fainter
population). The fact that a population of blue stars with ages $\la
100$ Myr is found in the companion \HI cloud, but no tracers of an
older population are found, provides evidence that star formation has
been triggered here only recently, very possibly for the first time.
In summary, our results show that the distribution of the atomic ISM
in dwarf irregular galaxies is not necessarily much more extended than
that of its stellar population.

\section*{Acknowledgments}
EdB acknowledges support from a PPARC Advanced Fellowship.

\clearpage

\end{document}